\documentclass[12pt]{iopart}

\usepackage{cite}
\usepackage[utf8]{inputenc}
\usepackage{graphicx}
\usepackage{graphicx,epstopdf}
\usepackage{dcolumn}
\usepackage{amsmath}
\usepackage{lineno,hyperref}
\usepackage{amsfonts}
\usepackage{amsthm}
\usepackage{bbm}
\usepackage{amssymb}
\usepackage{mathrsfs}
\PassOptionsToPackage{caption=false}{subfig}
\usepackage{subfig}
\usepackage{color}
\usepackage[normalem]{ulem}

\usepackage{mathtools}
\usepackage{array}
\usepackage{tabularx}  
\usepackage{booktabs}  
\usepackage{multirow}

\usepackage{blindtext}
\usepackage{lipsum}
\usepackage{bbm}

\usepackage{etoolbox}

\makeatletter
\newcommand{\mainmatter}{%
	\setcounter{footnote}{0}%
	\patchcmd{\@makefntext}{\fnsymbol}{\arabic}{}{}%
	\patchcmd{\@thefnmark}{\fnsymbol}{\arabic}{}{}%
	\def\@makefnmark{\textsuperscript{\arabic{footnote}}}%
}
\makeatother

\begin{document}

\title[]{Topological Classification of Dynamical Quantum Phase Transitions in the 1D XY model via Critical Mode Analysis}

\author{Bao-Ming Xu}
\address{Institute of Biophysics, Dezhou University, Dezhou 253023, China}
\ead{xubm2018@163.com}

\date{Submitted \today}

\begin{abstract}
Dynamical quantum phase transitions (DQPTs), which serve as a theoretical framework for understanding far-from-equilibrium physics in quantum many-body systems, have recently been observed experimentally. Their topological properties are typically characterized by the winding number, which acts as an order parameter. While DQPTs exhibiting both integer and half-integer jumps in the winding number have been reported, the underlying mechanisms behind these distinct topological behaviors, as well as the potential existence of other topological classes, remain open questions. To address this, we investigate DQPTs in the one-dimensional XY model under a quench protocol. We show that the observed topological diversity originates from the nature of the critical modes, which we classify into two categories: boundary modes and interior modes. Specifically, critical interior modes always lead to DQPTs with an integer winding number, while critical boundary modes always result in DQPTs characterized by a half-integer winding number. By analyzing the number and classification of critical modes, we provide a classification of the topological properties of DQPTs in the one-dimensional XY model. According to their distinct topological features, we categorize DQPTs into six types, three of which have not been previously identified in the literature. We discuss in detail the conditions associated with each type and present the corresponding dynamical phase diagrams. Our framework is not restricted to the XY model; it is applicable to other two-band models in one-dimensional systems, including the SSH model, Kitaev chain, Rice-Mele model, and Creutz model.
\end{abstract}
\vspace{2pc}

\maketitle

\section{Introduction}
The past decades have witnessed the flourishing of dynamical quantum phase transitions (DQPTs), both in experimental and theoretical research, among condensed matter physicists, because quantum simulators have nowadays achieved experimental access to the real-time dynamics of closed quantum many-body systems; see Refs. \cite{Heyl2018,Marino2022} for the recent reviews. Such quantum simulators have been realized on various experimental platforms, such as ultra-cold atoms \cite{Bloch2008,Greiner2002} or trapped ions \cite{Porras2004,Kim2009}. DQPTs are the sudden or non-analytic changes in the behaviour of a large quantum system during its time evolution, generalizing the nonanalytic behavior of the free energy at a phase transition in the thermodynamic limit to the out-of-equilibrium case \cite{Heyl2018,Heyl2013}. To this end, the Loschmidt echo, measuring the return probability of the time-evolving system onto its initial state, has been introduced. DQPT is defined by the zeros in the Loschmidt echo or the nonanalyticities in its rate function obtained by large deviation principle. It is analogous to the zeros in the partition function and the nonanalyticities in the free energy in equilibrium phase transitions, and therefore the rate function of Loschmidt echo is also called the dynamical free energy per particle.

Dynamical quantum phase transitions (DQPTs) were first introduced by Heyl et al. in the transverse-field Ising model \cite{Heyl2013}, where subsequent work established scaling and universality \cite{Heyl2015}. Extensions to other exactly solvable models, such as the Kitaev honeycomb \cite{Schmitt2015} and the Su-Schrieffer-Heeger model with flux quenches \cite{Rossi2022}, revealed how boundary conditions and lattice geometry affect dynamical criticality. Studies on nonintegrable systems \cite{Karrasch2013,Sharma2015} showed that DQPTs survive integrability breaking and can be controlled in long-range interacting spin models \cite{Kennes2018,Halimeh2021a}. In non-Hermitian systems, the Loschmidt echo acquires new features from complex spectra, linking non-Hermitian topology to dynamical phase transitions \cite{Zhou2018,Zhou2021a}. Subsequent work uncovered anomalous criticality in gapless non-Hermitian phases \cite{Mondal2022,Mondal2023}, culminating in biorthogonal DQPTs \cite{Jing2024}. The role of quantum state and temperature has also been explored. Extending to mixed states requires generalized Loschmidt echoes that account for decoherence \cite{Bhattacharya2017a,Heyl2017}, with a framework for open systems \cite{Lang2018} and finite-temperature generalizations \cite{Mera2018} probing thermal robustness. The driving protocol fundamentally shapes DQPTs: slow ramps reveal adiabatic-to-nonadiabatic crossovers \cite{Sharma2016}, quasiperiodic driving modifies entanglement and fidelity \cite{Maslowski2020}, and ramped quenches exhibit universal scaling \cite{Zamani2024,Cheng2025} as well as interplay with decoherence \cite{Baghran2024}. Periodically driven Floquet systems display distinct Floquet-DQPTs \cite{Yang2019,Zamani2020}, where periodicity controls topological transitions \cite{Jafari2022,Naji2022} and entanglement singularities \cite{Jafari2021}. Finally, disorder and noise modify critical exponents \cite{Trapin2021}, eliminate order parameters \cite{Kuliashov2023}, and compete with coherent dynamics \cite{Cao2020,Jafari2024}. Going beyond the equilibrium phase transition analogy, recent work has shown that the critical time obeys a power-law scaling with the quench rate, with the scaling exponent fully determined by the Kibble-Zurek mechanism \cite{Zhang2025}. Experimentally, DQPTs have been observed in diverse platforms, including trapped ions \cite{Jurcevic2017}, ultracold atoms \cite{Flaschner2018} and superconducting qubits \cite{Guo2019}.

In the theory of DQPT, the emergence of distinct phases is not immediately apparent from the nature of the effect, but the topological order parameters, such as the winding number \cite{Sharma2016,Budich2016,Bhattacharya2017b} and vortex loops \cite{Heyl2017,Kosior2024} have been identified as key indicators. The topological behavior of DQPT can be captured geometrically by the vector of momentum-wise Loschmidt overlap amplitude introduced by Ding \cite{Ding2020}. Depending on the trajectory of this vector, DQPTs are known to exhibit either integer \cite{Sharma2016,Budich2016,Bhattacharya2017b,Qiu2018} or half-integer \cite{Wong2023,Ding2020} jumps in the winding number. Nevertheless, the mechanisms underlying these distinct topological classifications, as well as the possible existence of other topological classes, remain largely unexplored. To address this gap, we investigate a one-dimensional XY model and classify the topological properties of DQPT through a detailed analysis of the critical mode. Specifically, we identify and characterize the critical modes $k^*$ by categorizing them into two types: critical boundary modes, where $k^\ast=0$ or $\pi$, and critical interior modes, where $k^{\ast}\in(0,\pi)$. We demonstrate that critical interior modes invariably give rise to DQPTs characterized by integer winding numbers, whereas critical boundary modes consistently yield DQPTs with half-integer winding numbers. By systematically analyzing the number and classification of these critical modes, we provide a framework for understanding the topological nature of DQPTs in the one-dimensional XY model. Our analysis reveals exactly six distinct types of topological behavior, three of which have not been previously identified in the literature.

This paper is organized as follows: In the next section, we introduce the quenched one-dimensional XY model, and briefly review the crucial method for the analytical solution, which will be used in the following. Sec. III is devoted to study the central objects of DQPT, including Loschmidt echo, the rate function, Fisher zeros and the topological order parameter of the winding number. In Sec. IV, a classification of DQPTs is provided, along with the associated criteria. According to the conditions under which different DQPTs occur, the dynamical phase diagrams are given in Sec. V. In Sec. VI, we discuss the other two-band mode in one-dimensional systems. Finally, Sec. VII closes the paper with some concluding remarks.

\begin{figure}
\begin{center}
\includegraphics[width=6cm]{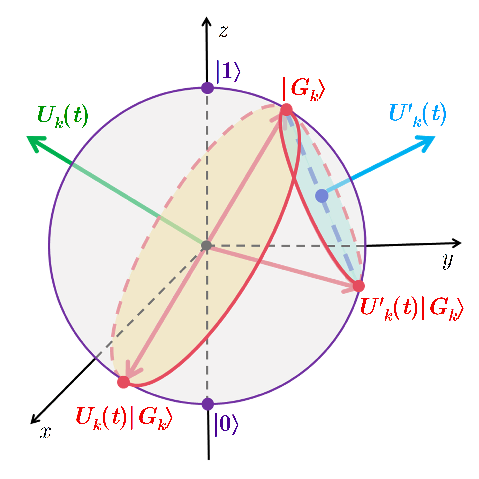}
\caption{(Color online) The schematic drawing of the Bloch sphere. In the Bloch sphere, $|1\rangle=\{\hat{c}^\dag_{-k}\hat{c}^\dag_k|0\rangle$ with $|0\rangle$ being the vacuum of the Jordan-Wigner fermions. $|1\rangle$ and $|0\rangle$ are the eigenstates of $\hat{\sigma}_z$, satisfying $\hat{\sigma}_z|1\rangle=|1\rangle$ and $\hat{\sigma}_z|0\rangle=-|0\rangle$.}
\label{figure1}
\end{center}
\end{figure}

\section{Quenched one-dimensional XY model}
The dynamical properties we are concerned with here are generated by one-dimensional XY model, whose Hamiltonian is
\begin{equation}\label{}
  \hat{H}(\lambda,\gamma)=-\frac{J}{2}\sum_{j=1}^{N}\biggl[\frac{1+\gamma}{2}\hat{\sigma}_j^x\hat{\sigma}_{j+1}^x
  +\frac{1-\gamma}{2}\hat{\sigma}_j^y\hat{\sigma}_{j+1}^y+\lambda\hat{\sigma}_j^z\biggr],
\end{equation}
where $J$ is longitudinal spin-spin coupling, $\gamma$ governs the anisotropic coupling between spins along the $x$ and $y$ directions, $\lambda$ is a dimensionless parameter measuring the strength of the transverse field with respect to the longitudinal spin-spin coupling. In this work, we set $J=1$ as the overall energy scale without loss of generality. $\hat{\sigma}^{\alpha}_{j}$ $(\alpha=x,y,z)$ is the spin-1/2 Pauli operator at lattice site $j$ and the periodic boundary conditions are imposed as $\hat{\sigma}^{\alpha}_{N+1}=\hat{\sigma}^{\alpha}_{1}$. Here we only consider that $N$ is even. The model exhibits competitions between anisotropic and magnetic couplings, which results in the existence of multiple phases. The quantum phase transition from the ferromagnetic phase (FM) to the paramagnetic phase (PM) driven by the transverse field $\lambda$ is called the Ising transition with the quantum critical point $\lambda_c=1$. On the other hand, the quantum phase transition between two FMs, with magnetic ordering in the $x$-direction and the $y$-direction, respectively, driven by the anisotropy parameter $\gamma$, is called the anisotropic transition with the critical point $\gamma_c=0$. In fact, in the absence of the transverse field, the ground state of the system is in the Luttinger liquid phase at $\gamma_c=0$.

The Hamiltonian is integrable and can be mapped to a system of free fermions and therefore be solved exactly. By applying the Jordan-Wigner transformation and the Fourier transformation, the Hamiltonian converts from spin operators into spinless fermionic operators as \cite{Bunder1999}
\begin{equation}\label{Hamiltonian}
\hat{H}(\lambda,\gamma)=\sum_k
\begin{pmatrix} \hat{c}^\dag_{-k} & \hat{c}_{k} \end{pmatrix}
\begin{pmatrix}\lambda+\cos k &-i\gamma\sin k \\ i\gamma\sin k & -\lambda-\cos k\end{pmatrix}
\begin{pmatrix} \hat{c}_{-k} \\ \hat{c}^\dag_{k} \end{pmatrix},
\end{equation}
where $\hat{c}_{k}$ and $\hat{c}^\dag_{k}$ are respectively fermion annihilation and creation operators for mode $k=(2n-1)\pi/N$ with $n=1\cdots N/2$, corresponding to periodic boundary conditions for $N$ is even. Each $\hat{H}_k(\lambda,\gamma)$ acts on a two-dimensional Hilbert space generated by $\{\hat{c}^\dag_{-k}\hat{c}^\dag_k|0\rangle,~|0\rangle\}$, where $|0\rangle$ is the vacuum of the Jordan-Wigner fermions $\hat{c}_k$ and $\hat{c}_{-k}$, and can be represented in that basis by a $2\times2$ matrix
\begin{equation}\label{}
  \hat{H}_k(\lambda,\gamma)=\mathbf{d}_k\cdot\hat{\mathbf{\sigma}}
\end{equation}
with
\begin{equation}\label{dk}
  \mathbf{d}_k=(0,\gamma\sin k,\lambda+\cos k).
\end{equation}
The instantaneous eigenvalues are $d^{\pm}_k=\pm d_{k}$ with
\begin{equation}\label{}
d_{k}=\sqrt{(\lambda+\cos k)^2+\gamma^2\sin^2k}.
\end{equation}
The corresponding eigenvectors are
\begin{equation}\label{}
|d^+_{k}\rangle=
\biggl[\frac{1+d_k^z/d_k}{\sqrt{2(1+d_k^z/d_k)}}\hat{c}^\dag_{-k}\hat{c}^\dag_k+i\frac{d_k^y/d_k}{\sqrt{2(1+d_k^z/d_k)}}\biggr]|0\rangle
\end{equation}
and
\begin{equation}\label{ground state}
|d^-_{k}\rangle=
\biggl[i\frac{d_k^y/d_k}{\sqrt{2(1+d_k^z/d_k)}}\hat{c}^\dag_{-k}\hat{c}^\dag_k+\frac{1+d_k^z/d_k}{\sqrt{2(1+d_k^z/d_k)}}\biggr]|0\rangle,
\end{equation}
respectively.

At time $t\leq0$, the system is prepared in the ground state
\begin{equation}\label{}
  |G\rangle=\bigotimes_k|d^-_{k}\rangle.
\end{equation}
Then at time $t=0$, the transverse field is suddenly changed from $\lambda$ to $\lambda'$ or the anisotropic parameter is suddenly changed from $\gamma$ to $\gamma'$. We assume that this process is so sudden that the system state has no time to change. After quenching, the dynamics of the system is governed by the evolution operator ($\hbar=1$)
\begin{equation}\label{U}
  \hat{U}(t)=e^{-i\hat{H}(\lambda',\gamma')t}=\bigotimes_ke^{-i\hat{H}_k(\lambda',\gamma')t},
\end{equation}
so that the state $|\psi(t)\rangle$ at a time $t$ after the quench is given by
\begin{equation}\label{}
 |\psi(t)\rangle=\bigotimes_ke^{-i\hat{H}_k(\lambda',\gamma')t}|d^-_{k}\rangle.
\end{equation}

\section{The fundamental theory of DQPT}

The central object within the theory of DQPTs is the Loschmidt overlap amplitude quantifying the deviation of the time-evolved state from the initial condition. The Loschmidt overlap amplitude is defined as
\begin{equation}\label{overlap1}
  \mathcal{G}(t)=\langle G|\hat{U}(t)|G\rangle.
\end{equation}
Because different modes are independent with each other, the Loschmidt overlap amplitude can be written as
\begin{equation}\label{overlap2}
  \mathcal{G}(t)=\prod_k\mathcal{G}_k(t)
  =\prod_k\langle d^-_{k}|e^{-i\hat{H}_k(\lambda',\gamma')t}|d^-_{k}\rangle.
\end{equation}
The rate function of the Loschmidt echo, measuring the return probability $|\mathcal{G}(t)|^2$, is defined as
\begin{equation}\label{}
r(t)=-\lim_{N\rightarrow\infty}\frac{1}{N}\ln|\mathcal{G}(t)|^2.
\end{equation}
It is the dynamical analogue of the equilibrium free energy, namely the dynamical free energy. In the thermodynamic limit one can derive an exact result for $r(t)$:
\begin{equation}\label{}
  r(t)=-\frac{1}{\pi}\mathrm{Re}\biggl[\int_0^\pi dk \ln\mathcal{G}_k(t)\biggr].
\end{equation}
A detailed derivation of the rate function can be found in the appendix.
A DQPT will occur when $r(t)$ exhibits a nonanalyticity at some times $t^*$, namely critical times.

A powerful method to analyse the nonanalyticity of $r(t)$ is Fisher zeros \cite{Heyl2013,Fisher1965,Cheraghi2024}. Fisher zeros provide a fundamental basis for determining whether DQPT occurs, and have been extensively used to explore the connection between DQPTs and equilibrium phase transitions \cite{Vajna2014}. In order to apply this concept, we should expend time $t$ into the complex plane and focus on the boundary partition function $Z(z)=\langle G|e^{-z\hat{H}(\lambda',\gamma')}|G\rangle $. The boundary partition function can be expressed as
\begin{equation}\label{Zz}
  Z(z)=\prod_k\langle d^-_{k}|e^{-z\hat{H}_k(\lambda',\gamma')}|d^-_{k}\rangle
\end{equation}
where $z\in\mathbb{C}$. For imaginary $z=it$ this just describes the overlap amplitude of Eq. (\ref{overlap1}). The Fisher zeros are the values of $z$ that make $Z(z)=0$. Due to the multiplicative properties of Eq. (\ref{Zz}), a zero in $Z(z)$ is equivalent to finding at least one mode $k$ and one $z$ making $\langle d^-_{k}|e^{-z\hat{H}_{k}(\lambda',\gamma')}|d^-_{k}\rangle = 0$. In the thermodynamic limit the zeros of the boundary partition function in the complex plane coalesce to a family of lines labeled by a number $n\in\mathbb{Z}$
\begin{equation}\label{Zeros}
z_n(k)=\frac{1}{2d_{k}'}\Biggl[\ln\frac{1-\frac{\mathbf{d}_k}{d_k}\cdot\frac{\mathbf{d}_k'}{d_k'}}
{1+\frac{\mathbf{d}_k}{d_k}\cdot\frac{\mathbf{d}_k'}{d_k'}}+i(2n+1)\pi\Biggr].
\end{equation}
A detailed derivation of the Fisher zeros can be found in the appendix.
DQPTs, manifested as the non-analytic behaviors of the dynamical free energy, are contingent upon the existence of purely imaginary zeros of the boundary partition function, corresponding to real-time zeros of the Loschmidt overlap amplitude. From Eq. (\ref{Zeros}), we can obtain the orthogonality condition of DQPT \cite{Ding2020,Wong2023}
\begin{equation}\label{orthogonal condition}
\mathbf{d}_k\cdot\mathbf{d}_k'=0.
\end{equation}
In other words, given a protocol, a DQPT can only occur if at least one mode satisfy Eq. \eqref{orthogonal condition}, this mode is called critical mode $k^*$. Then, there is a series of critical times $t^*_n$
\begin{equation}\label{critical time}
  t^*_n=\frac{1}{2d_{k^*}'}(2n+1)\pi
\end{equation}
depending on the corresponding critical mode $k^*$.

Physically, the orthogonality condition of DQPT can be understood with the help of the Bloch sphere. According to the rate function, DQPT occurring is equivalent to making $\langle d^-_{k^*}|e^{-i\hat{H}_{k^*}(\lambda',\gamma')t}|d^-_{k^*}\rangle = 0$, which means a flip of the vector of $|d^-_{k^*}\rangle$ in the Bloch sphere (see Fig. \ref{figure1}). It is conceivable that the axis of rotation must be perpendicular to a vector in order to flip it (see the green axis in Fig. \ref{figure1}). Therefore, DQPT is only possible if the axis corresponding to the evolution operator $\hat{U}_{k^*}(t)=e^{-i\hat{H}_{k^*}(\lambda',\gamma')t}$ is perpendicular to the initial state $|d^-_{k^*}\rangle$, otherwise it is impossible (see the blue axis in Fig. \ref{figure1}). The axis corresponding to the evolution operator can be understood as $\mathbf{d}_{k^*}'$ and the vector of initial state can be understood as $\mathbf{d}_{k^*}$, therefore the condition of DQPT can be demonstrated as $\mathbf{d}_{k^*}\cdot\mathbf{d}_{k^*}'=0$.

However, it is important to note that the condition $\mathbf{d}_{k^*}\cdot\mathbf{d}_{k^*}'=0$ is not always sufficient to induce DQPT. Specifically, the preceding discussions have implicitly assumed $d_{k^*}\neq0$ and $d_{k^*}'\neq0$, i.e., the energy levels of the pre- and post-quench systems are both non-degenerated. If this assumption is violated, i.e., $d_{k^*}=0$ or $d_{k^*}'=0$, the condition $\mathbf{d}_{k^*}\cdot\mathbf{d}_{k^*}'=0$ no longer guarantees perpendicularity. This is because the orientation of zero vector is undefined and can not be used to represent the orientation of the initial state or the axis of evolution. To address this subtlety, we use a more fundamental and robust criterion for determining DQPT: Whether the initial state can be flipped. If $d_{k^*}'=0$, the initial state $|d^-_{k^*}\rangle$ would not evolve at all, let along undergo DQPT, because $\hat{U}_{k^*}(t)=e^{-i\hat{H}_{k^*}(\lambda',\gamma')t}=\mathbb{I}$ ($\mathbb{I}$ is an identity matrix). If $d_{k^*}=0$ but $d_{k^*}'\neq0$, the orientation of the evolution operator would still be represented by $\mathbf{d}_{k^*}'$, and as long as it is perpendicular to the orientation of the initial state (not $\mathbf{d}_{k^*}$), the initial state would be flipped and DQPT would occur. The analysis above will be concretized in the following sections.

\begin{table}[]
\centering
\caption{The conditions for DQPTs. The conditions for the occurrence of DQPTs involve four key components: the relationship between the anisotropy parameters, denoted by $\gamma\rightarrow\gamma'$; the relationship between the transverse fields, denoted by $\lambda\rightarrow\lambda'$; the quantity $\Delta$ (defined in Eq.~\eqref{Delta}); and the expression $\frac{\gamma\gamma'+\lambda\lambda'}{1-\gamma\gamma'}$. A given type of DQPT occurs precisely when the constraints associated with all four components--aligned in the same row of the table--are simultaneously satisfied. Some types of DQPT can arise from multiple independent sets of conditions. For instance, DQPT-1 can be realized under four independent scenarios, each corresponding to a distinct row in the table. In cases where a given cell contains only a single constraint, it indicates that the corresponding component is subject solely to that condition. As an example, for DQPT-4, the component concerning the transverse field relationship (i.e., ``$\lambda\rightarrow\lambda'$") is governed exclusively by the constraint $|\lambda|=1$, $\lambda'\neq\lambda$. This means that although three independent condition sets can lead to DQPT-4, they all share the same requirement for the transverse field relationship (i.e., ``$\lambda\rightarrow\lambda'$"). The symbol ``---" in the table indicates that the corresponding component does not impose any constraint.}
\begin{tabular}{c|c|c|c|c}
\hline\hline
\multirow{2}{*}{DQPTs} & \multicolumn{4}{c}{conditions}            \\ \cline{2-5}
& $\gamma\rightarrow\gamma'$ &$\lambda\rightarrow\lambda'$ &$\Delta$ &$\frac{\gamma\gamma'+\lambda\lambda'}{1-\gamma\gamma'}$\\
\hline
\multirow{4}{*}{DQPT-1} & $\gamma\gamma'\neq0 $      & $(|\lambda|-1)(|\lambda'|-1)<0$        &\multirow{4}{*}{---} & ---     \\
                        & $\gamma=0$, $\gamma'\neq0$ & $|\lambda'|<1$                         &  & ---     \\
                        & $\gamma\neq0$, $\gamma'=0$ & $|\lambda|<1$, $\lambda'\neq\lambda$   &  & ---     \\
                        & $\gamma\gamma'(\gamma\gamma'-1)\neq0$       & $\lambda\neq\lambda'$, $|\lambda'|=1$  &  & $<1$    \\
\hline
               DQPT-2   & $\gamma\gamma'(\gamma\gamma'-1)\neq0$      & $(|\lambda|-1)(|\lambda'|-1)>0$        & $>0$  & $<1$  \\
\hline
               DQPT-3   & $\gamma\gamma'(\gamma\gamma'-1)\neq0$       & ---        & =0  & $<1$ \\
\hline
\multirow{3}{*}{DQPT-4} & $\gamma\gamma'=1$          &\multirow{3}{*}{$|\lambda|=1$, $\lambda'\neq\lambda$}   & --- & ---   \\
                         & $\gamma\neq0$, $\gamma'=0$ &                                                      & --- & ---   \\
                        & $\gamma\gamma'(\gamma\gamma'-1)\neq0$       &                             & $>0$  & $>1$  \\
\hline
               DQPT-5   & $\gamma\gamma'(\gamma\gamma'-1)\neq0$  & $|\lambda|=1$, $\lambda'\neq\lambda$                   & $>0$ & $<1$  \\
\hline
               DQPT-6   & $\gamma\gamma'=1$          & $|\lambda|=1$, $\lambda'=-\lambda$                      & --- & ---   \\

\hline\hline
\end{tabular}
\label{Table}
\end{table}

Beyond being characterized by non-analytic singularities in the rate function, DQPTs lack a clear phase classification, in contrast to their equilibrium or quantum phase transition counterparts. To address this limitation, a topological classification based on the winding number was proposed \cite{Sharma2016,Budich2016,Bhattacharya2017b}.
At a given time, the winding number of is defined as
\begin{equation}\label{}
  \nu(t)=\frac{1}{2\pi}\int_0^\pi\frac{\partial\phi_k^G(t)}{\partial k}dk,
\end{equation}
where
\begin{equation}\label{}
  \phi_k^G(t)=\phi_k(t)-\phi_k^D(t)
\end{equation}
is the geometric phase of the Loschmidt overlap amplitude, with
\begin{equation}\label{}
  \phi_k(t)=\arg\mathcal{G}_k(t)
\end{equation}
being the total phase of $\mathcal{G}_k(t)$ and
\begin{equation}\label{}
  \phi_k^D(t)=-\langle d^-_{k}|\hat{H}_k(\gamma',\lambda')|d^-_{k}\rangle t.
\end{equation}
being the dynamical phase. The behavior of the winding number can be visualized by the vector of the momentum-resolved Loschmidt overlap amplitude introduced by Ding \cite{Ding2020}:
\begin{equation}\label{trajectory}
\vec{R}_k = (x_k, y_k)= |\mathcal{G}_k(t)|e^{i\phi_k^G(t)}.
\end{equation}
In this representation, the winding number corresponds precisely to the number of times the trajectory $\vec{R}_k$ winds around the origin. Typically, whenever the system evolves through a critical time of a DQPT, the winding number exhibits a quantized jump, taking either integer \cite{Sharma2016,Budich2016,Bhattacharya2017b,Qiu2018} or half-integer \cite{Wong2023,Ding2020} values; it is therefore commonly regarded as a dynamical topological order parameter for DQPTs. Nevertheless, the mechanisms underlying these distinct topological classifications, as well as the possible existence of other topological classes, remain largely unexplored. To address this gap is the main goal of this paper.

\begin{figure*}
\begin{center}
\includegraphics[width=14cm]{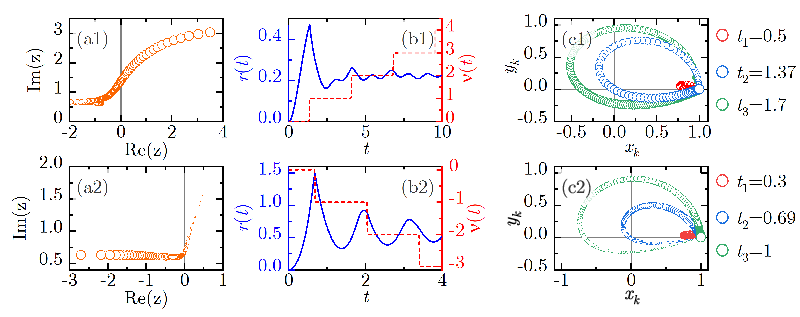}
\caption{(Color online) The properties of DQPT-1 are illustrated by the lines of Fisher zeros (a1, a2), the time evolution of the rate function $r(t)$ (blue curves in b1 and b2) and the winding number $\nu(t)$ (red curves in b1 and b2), as well as and the trajectory of vector $\vec{R}_k=(x_k,y_k)$ (c1 and c2) for the quenches from $\lambda=0.5$ to $\lambda'=1.5$ (a1-c1) and $\lambda'=-1.5$ (a2-c2). Trajectories are plotted before (red bubbles), at (blue bubbles) and after (green bubbles) the first critical time $t_1^*$. The other parameters are $\gamma=0.5$, $\gamma'=2$ and $n=0$.}
\label{figure2}
\end{center}
\end{figure*}

\begin{figure*}
\begin{center}
\includegraphics[width=14cm]{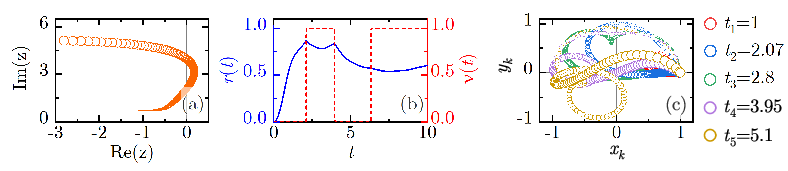}
\caption{(Color online) The properties of DQPT-2 are illustrated by the lines of Fisher zeros (a), the time evolution of rate function $r(t)$ (blue curves in b), the winding number $\nu(t)$ (red curves in b) and the trajectory of vector $\vec{R}_k=(x_k,y_k)$ (c) for the quenches from $\lambda=1.1$ to $\lambda'=1.3$. The first and the second critical times are $t_1^*\approx2.07$ and $t_2^*\approx3.95$. The other parameters are $\gamma=1$, $\gamma'=-0.5$ and $n=0$.}
\label{figure3}
\end{center}
\end{figure*}

\section{The topological classification of DQPTs}

In this section, we provide a topological classification of DQPTs in one-dimensional XY models and map out the corresponding parameter space in which these phenomena occur. Our approach is primarily based on a detailed analysis of the critical mode(s). Specifically, we designate the modes at $k=0$ and $k=\pi$ as boundary modes, while the remaining modes, i.e., $k\in(0,\pi)$, are referred to as interior modes. A critical mode can be either a boundary or an interior mode. Based on the number and nature of the critical modes, together with the properties of Fisher zeros, we propose a framework for understanding the topological properties of DQPTs. In previous works \cite{Vajna2015,Porta2020}, DQPTs in two-band models have been classified according to the ground-state topology of both the initial and final Hamiltonians. In contrast, our investigation is based on the type and number of critical modes, with each resulting DQPT exhibiting distinct topological characteristics.

The critical mode is obtained by solving the orthogonality condition of Eq. \eqref{orthogonal condition}. According to Eq.~\eqref{dk}, the orthogonality condition takes the form
\begin{equation}\label{critical mode}
(\lambda+\cos k)(\lambda'+\cos k)+\gamma\gamma'\sin^2 k=0.
\end{equation}
Solving this equation yields the critical modes $k^*$; the conditions under which such solutions exist determine the parameter sets $(\lambda,\gamma,\lambda',\gamma')$ for which DQPTs appear. By analyzing the number and type of these critical modes, we further characterize the nature of the associated DQPT.

All subsequent properties and discussions follow from solving Eq.~\eqref{critical mode}; therefore, we first present its solutions.

When $\gamma\gamma'=0$, Eq. (\ref{critical mode}) yields the solution
\begin{equation}\label{}
  \cos k=-\lambda, ~ \mathrm{or} ~ -\lambda'.
\end{equation}

If $\gamma\gamma'=1$, the solution is
\begin{equation}\label{solution1}
  \cos k=-\frac{1+\lambda\lambda'}{\lambda+\lambda'}
\end{equation}
provided that $\lambda+\lambda'\neq0$.

If $\gamma\gamma'(\gamma\gamma'-1)\neq0$, the solutions of Eq. (\ref{critical mode}) are
\begin{equation}\label{solution2}
  \cos k=\frac{1}{2(1-\gamma\gamma')}\Bigl[-(\lambda+\lambda')\pm\sqrt{\Delta}\Bigr],
\end{equation}
where $\Delta$ is defined as
\begin{equation}\label{Delta}
\Delta=(\lambda+\lambda')^2-4(1-\gamma\gamma')(\gamma\gamma'+\lambda\lambda')\geq0.
\end{equation}

The above solutions and the conditions for their existence can be summarized as follows:
\begin{equation} \label{solutions}
\cos k=
\begin{cases}
-\lambda, ~ \mathrm{or} ~ -\lambda' & \text{if $\gamma\gamma'=0$}, \\
-\frac{1+\lambda\lambda'}{\lambda+\lambda'} & \text{if $\gamma\gamma'=1$}, \\
\frac{1}{2(1-\gamma\gamma')}\Bigl[-(\lambda+\lambda')\pm\sqrt{\Delta}\Bigr] & \text{if $\gamma\gamma'(\gamma\gamma'-1)\neq0$ and $\Delta\geq0$}.
\end{cases}
\end{equation}
In the following, to simplify the notation, we denote the expression $\frac{1}{2(1-\gamma\gamma')}\Bigl[-(\lambda+\lambda')\pm\sqrt{\Delta}\Bigr]$ as $F_{\pm}$, i.e.,
\begin{equation}\label{FF}
F_{\pm}=\frac{1}{2(1-\gamma\gamma')}\Bigl[-(\lambda+\lambda')\pm\sqrt{\Delta}\Bigr],
\end{equation}
as it will appear repeatedly throughout the discussion.

It is worth noting that these solutions are all formal solutions; whether critical modes exist depends on the specific values of the parameters. In what follows, we will carefully examine the conditions for the existence of critical modes, analyze their nature and number in detail, and determine the corresponding parameter spaces. Based on the properties and number of critical modes, the topological characteristics of DQPT are discussed, and these transitions are classified according to their distinct topological properties.

\subsection{DQPT-1}
We first discuss the case of a single critical interior mode, where the line of Fisher zeros cuts $\mathrm{Im}(z)$ axis. For convenience, we denote the DQPT that occurs in this case as DQPT-1. Below, we discuss the parameter space and topological properties of DQPT-1 in detail, based on the solutions in Eq.~\eqref{solutions}.

\textbf{Parameter Space of DQPT-1}

1. $\gamma\gamma'=1$:

A critical interior mode $k^*=\arccos[-\frac{1+\lambda\lambda'}{\lambda+\lambda'}]$ can be found if $|\frac{1+\lambda\lambda'}{\lambda+\lambda'}|<1$. And we find that $(\mathbf{d}_{k<k^*}\cdot\mathbf{d}_{k<k^*}')(\mathbf{d}_{k>k^*}\cdot\mathbf{d}_{k>k^*}')<0$, which means that the line of Fisher zeros cuts $\mathrm{Im}(z)$ axis at this critical interior mode, thereby implying the occurrence of DQPT-1. The condition $|\frac{1+\lambda\lambda'}{\lambda+\lambda'}|<1$ is equivalent to
\begin{equation}\label{condition1}
(|\lambda|-1)(|\lambda'|-1)<0.
\end{equation}
In other words, when the transverse field is quenched across the quantum critical point $|\lambda_c|=1$, i.e., from $|\lambda|<1$ to $|\lambda'|>1$ or vice versa, DQPT-1 occurs.

\begin{figure*}
\begin{center}
\includegraphics[width=14cm]{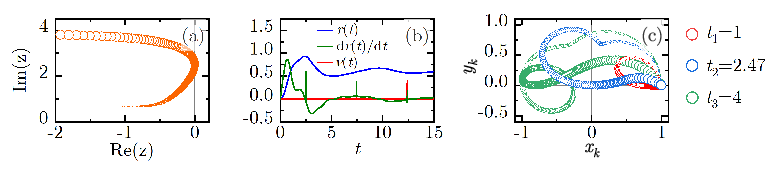}
\caption{(Color online) The properties of DQPT-3 are illustrated by the lines of Fisher zeros (a), the time evolution of rate function $r(t)$ (blue curve in b), its derivative $\mathrm{d}r(t)/\mathrm{d}t$ (green curve in b) and the winding number $\nu(t)$ (red curve in b), as well as the trajectory of vector $\vec{R}_k=(x_k,y_k)$ (c). The parameters are $\lambda=1.1$, $\lambda'=(1-2\gamma\gamma')\lambda-\sqrt{4\gamma\gamma'(1-\gamma\gamma')(1-\lambda^2)}$. The first critical times is $t_1^*\approx2.47$. The other parameters are $\gamma=1$, $\gamma'=-0.5$ and $n=0$.}
\label{figure4}
\end{center}
\end{figure*}

\begin{figure*}
\begin{center}
\includegraphics[width=14cm]{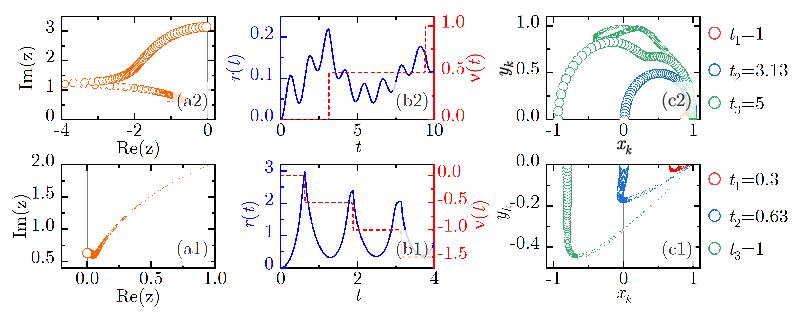}
\caption{(Color online) The properties of DQPT-4 are illustrated by the lines of Fisher zeros (a1 and a2), the time evolution of rate function $r(t)$ (blue curves in b1 and b2), the winding number $\nu(t)$ (red curves in b1 and b2) and the trajectory of vector $\vec{R}_k=(x_k,y_k)$ (c1 and c2) for the quenches from $\lambda=1$ to $\lambda'=1.5$ (a1-c1) and $\lambda'=-1.5$ (a2-c2). Trajectories are plotted before (red bubbles), at (blue bubbles) and after (green bubbles) the first critical time $t_1^*$. The other parameters are $\gamma=0.5$, $\gamma'=2$ and $n=0$.}
\label{figure5}
\end{center}
\end{figure*}

2. $\gamma\gamma'=0$:

Two critical interior modes $k^*_1=\arccos(-\lambda)$ and $k^{*}_2=\arccos(-\lambda')$ could be found if $|\lambda|<1$ and $|\lambda'|<1$. However, whether DQPT-1 occurs depends on the quench protocol of anisotropic parameter. Specifically, if the quench is from $\gamma=0$ to $\gamma'\neq0$, the initial state $|d_{k}^-\rangle=|0\rangle$ can be flipped by the evolution operator $\hat{U}_{k^{*}_2}(t)=e^{-id_{k^{*}_2}'\sigma_yt}$ but not by the evolution operator $\hat{U}_{k^*_1}(t)=e^{-i\hat{H}_{k^*_1}(\lambda',\gamma')t}$, therefore DQPT only occurs at the critical interior mode $k^{*}_2=\arccos(-\lambda')$, where the line of Fisher zeros cuts $\mathrm{Im}(z)$ axis at this critical mode. Conversely, for a quench from $\gamma\neq0$ to $\gamma'=0$, DQPT-1 would occur at the critical mode $k^*_1=\arccos(-\lambda)$, but not at the critical mode $k^{*}_2=\arccos(-\lambda')$, since $d_{k^*_2}'=0$ prevents the initial state from evolving at all. In summary, when $\gamma\gamma'=0$, DQPT-1 occurs under the following conditions: (i) $\gamma=0\rightarrow\gamma'\neq0$ with $|\lambda'|<1$, or (ii) $\gamma\neq0\rightarrow\gamma'=0$ with $|\lambda|<1$ and $\lambda\neq\lambda'$.

3. $\gamma\gamma'(\gamma\gamma'-1)\neq0$ and $\Delta>0$:

According to solutions of Eqs. \eqref{solutions} and \eqref{FF}, a single critical interior mode $k^*=\arccos F_+$ could be found if $|F_+|<1$ and $|F_-|>1$, indicating the occurrence of DQPT-1. Or conversely, a single interior critical mode $k^*=\arccos F_-$ could be found if $|F_-|<1$ and $|F_+|>1$. This condition of $|F_\pm|<1$ and $|F_\mp|>1$ can be expressed as $(F_+-1)(F_--1)<0$, which is equivalent to $(|\lambda|-1)(|\lambda'|-1)<0$, meaning that the transverse field is quenched crossing the quantum critical point $|\lambda_c|=1$. It can be proven that $(|\lambda|-1)(|\lambda'|-1)<0$ implies $\Delta>0$, therefore only $(|\lambda|-1)(|\lambda'|-1)<0$ deserves consideration in the parameter space of DQPT-1.

We note that the solutions of Eq. \eqref{solution2} reduce to $\cos k_1=\mp1$ and $\cos k_2=\mp\frac{\lambda\lambda'+\gamma\gamma'}{1-\gamma\gamma'}$ when either $\lambda=\pm1$ or $\lambda'=\pm1$. Consequently, two critical modes, $k^*_1=\pi$ or $0$ and $k^*_2=\arccos\bigl(\mp\frac{\lambda\lambda'+\gamma\gamma'}{1-\gamma\gamma'}\bigr)$, can be found provided that
\begin{equation}\label{c3}
\Bigl|\frac{\lambda\lambda'+\gamma\gamma'}{1-\gamma\gamma'}\Bigr|<1.
\end{equation}
In order to observe DQPT-1, the effects of the critical boundary mode $k^*_1=\pi$ or $0$ must be suppressed. To achieve this, the transverse field should be quenched to the critical point $\lambda'=\pm1$, so that the initial state for the critical boundary mode does not evolve at all. In this case, the critical interior mode $k^*_2=\arccos-\frac{\lambda\lambda'+\gamma\gamma'}{1-\gamma\gamma'}$ ($\lambda'=1$) or $k^*_2=\arccos\frac{\lambda\lambda'+\gamma\gamma'}{1-\gamma\gamma'}$ ($\lambda'=-1$) cause DQPT-1. In summary, when $\gamma\gamma'(\gamma\gamma'-1)\neq0$, DQPT-1 occurs under the following conditions: (i) $(|\lambda|-1)(|\lambda'|-1)<0$, or (ii) $|\frac{\lambda\lambda'+\gamma\gamma'}{1-\gamma\gamma'}|<1$ with $|\lambda'|=1$ and $\lambda\neq\lambda'$.

All the elements consisting the parameter space of DQPT-1 are investigated above. To sum up, they consist of the following cases (see Tab. \ref{Table}):

(i) $\gamma\gamma'\neq0$, $(|\lambda|-1)(|\lambda'|-1)<0$;

(ii) $\gamma=0$, $\gamma'\neq0$, $|\lambda'|<1$;

(iii) $\gamma\neq0$, $\gamma'=0$, $|\lambda|<1$, $\lambda'\neq\lambda$;

(iv) $\gamma\gamma'(\gamma\gamma'-1)\neq0$, $\Delta>0$, $\lambda\neq\lambda'$, $|\lambda'|=1$, $|\frac{\lambda\lambda'+\gamma\gamma'}{1-\gamma\gamma'}|<1$.

\textbf{Topological properties of DQPT-1}

To illustrate the properties of DQPT-1, we consider $\gamma=0.5\rightarrow\gamma'=2$ and $\lambda=0.5\rightarrow\lambda'=\pm1.5$ as an example. Fig. \ref{figure2} plots the lines of Fisher zeros, the time evolution of rate function $r(t)$, the winding number $\nu(t)$ and the trajectory of vector $\vec{R}_k=(x_k,y_k)$. It can be seen that the line of Fisher zeros cuts $\mathrm{Im}(z)$ axis at the critical mode [see Figs. \ref{figure2}(a1) and (a2)], giving rise to nonanalytic behavior (cusp singularity) of the rate function of the Loschmidt echo [see the blue curves in Figs. \ref{figure2}(b1) and (b2)]. The trajectory of vector $\vec{R}_k$ forms a closed loop anticounterclockwise [see Fig. \ref{figure2}(c1)] or counterclockwise for $\lambda'<0$ [see Fig. \ref{figure2}(c2)], as $k$ varies from $0$ to $\pi$. The trajectory does not encircle the origin until it firstly crosses it at the first critical time [see Figs. \ref{figure2}(c1) and (c2)], and then the winding number jumps up into $1$ [see the red curves in Fig. \ref{figure2}(b1)] or jumps down into $-1$ [see the red curves in Fig. \ref{figure2}(b2)]. After the first critical time, the trajectory will encircle the origin until the second critical time, when the trajectory secondly crosses the origin and the winding number jumps up into 2 or jumps down into $-2$, and so on. It should be noted that DQPT-1 corresponds to the scenario first reported in the seminal work by Heyl \cite{Heyl2013}.

\subsection{DQPT-2}

Here, we discuss the case of two critical interior modes. For convenience, we denote the DQPT that occurs in this case as DQPT-2. Below, we discuss the parameter space and topological properties of DQPT-2 in detail, based on the solutions in Eq. \eqref{solutions}. According to Eqs. \eqref{solutions} and \eqref{FF}, these two critical interior modes can only be $k^*_{+}=\arccos F_+$ and $k^*_{-}=\arccos F_-$.

\textbf{Parameter Space of DQPT-2}

From the solutions of Eqs. \eqref{solutions} and \eqref{FF}, it can be seen that the presence of two critical interior modes requires $(|F_+|-1)(|F_-|-1)>0$ and $|F_+F_-|<1$. $(|F_+|-1)(|F_-|-1)>0$ implies that \begin{equation}\label{same phase}
(|\lambda|-1)(|\lambda'|-1)>0,
\end{equation}
i.e., the transverse field is quenched within the same phases. $|F_+F_-|<1$ is equivalent to $|\frac{\gamma\gamma'+\lambda\lambda'}{1-\gamma\gamma'}|<1$ [Eq. (\ref{c3})]. Together with the solution existence conditions $\gamma\gamma'(\gamma\gamma'-1)\neq0$ and $\Delta>0$ (see Eq. \eqref{solutions}), these constraints define the parameter space for DQPT-2, which can be summarized as $\gamma\gamma'(\gamma\gamma'-1)\neq0$, $(|\lambda|-1)(|\lambda'|-1)>0$, $\Delta>0$ and $|\frac{\gamma\gamma'+\lambda\lambda'}{1-\gamma\gamma'}|<1$,  as list in Tab. \ref{Table}.

\textbf{Topological properties of DQPT-2}

Taking $\gamma=1\rightarrow\gamma'=-0.5$ and $\lambda=1.1\rightarrow\lambda'=1.3$ as an example, we observe that the line of Fisher zeros cuts $\mathrm{Im}(z)$ axis twice at two critical modes [see Fig. \ref{figure3}(a)], giving rise to nonanalytic behavior (cusp singularity) of the rate function which implies DQPT occurring [see the blue curves in Fig. \ref{figure3}(b)]. The trajectory of vector $\vec{R}_k$ forms a closed loop anticounterclockwise as $k$ varies from $0$ to $\pi$. The trajectory does not encircle the origin until it firstly crosses it at the first critical time $t_1^*\approx2.07$ [see Fig. \ref{figure3}(c)], and then the winding number jumps up into $1$ [see the red curves in Fig. \ref{figure3}(b)]. After the first critical time, the trajectory will encircle the origin until the second critical time $t_2^*\approx3.95$, when it puts the origin out the loop, resulting in the winding number jumping down to 0. DQPT-2 has been discussed in Ref. \cite{Vajna2014,Budich2016}, and has been shown to occur with the aid of quantum coherence \cite{Xu2024}.

Taking $\gamma=1\rightarrow\gamma'=-0.5$ and $\lambda=1.1\rightarrow\lambda'=1.3$ as an example, we observe that the line of Fisher zeros intersects the $\mathrm{Im}(z)$ axis twice, corresponding to two distinct critical modes [see Fig. \ref{figure3}(a)]. This gives rise to nonanalytic behavior (cusp singularities) in the rate function, signaling the occurrence of DQPTs [see the blue curve in Fig. \ref{figure3}(b)]. As $k$ varies from $0$ to $\pi$, the trajectory of $\vec{R}_k$ forms a closed loop that winds counterclockwise. Initially, the trajectory does not enclose the origin until it first crosses it at the first critical time $t_1\approx2.07$ [see Fig. \ref{figure3}(c)], at which point the winding number jumps to $1$ [see the red curve in Fig. \ref{figure3}(b)]. After this first critical time, the trajectory encircles the origin until the second critical time $t_2\approx3.95$, when the origin is no longer enclosed, causing the winding number to drop back to $0$. This type of DQPT, referred to as DQPT-2 corresponds to phenomena previously observed in Refs. \cite{Vajna2014,Budich2016} and has been linked to quantum coherence effects \cite{Xu2024}.

\subsection{DQPT-3}

Notably, two Fisher zeros cutting $\mathrm{Im}(z)$ axis [see Fig. \ref{figure3}(a)] can approach each other, merge, and subsequently move away from the $\mathrm{Im}(z)$ axis, by adjusting the external field and anisotropic parameter. The merged Fisher zero indicates that there is only one solution to Eq. (\ref{critical mode}), given by $F_+=F_-=-\frac{\lambda+\lambda'}{2(1-\gamma\gamma')}$ [see Eqs. \eqref{solutions} and \eqref{FF}], where the condition $\Delta=0$ is used. This solution corresponds to a critical interior mode $k^*=\arccos[-\frac{\lambda+\lambda'}{2(1-\gamma\gamma')}]$. At this critical mode, $(\mathbf{d}_{k<k^*}\cdot\mathbf{d}_{k<k^*}')(\mathbf{d}_{k>k^*}\cdot\mathbf{d}_{k>k^*}')>0$, which implies that the line of Fisher zeros only touches but not cross $\mathrm{Im}(z)$ axis. The DQPT arising from this unique critical interior mode is termed DQPT-3, representing an entirely new class of DQPT.

\textbf{Parameter Space of DQPT-3}

The presence of the critical mode $k^*=\arccos[-\frac{\lambda+\lambda'}{2(1-\gamma\gamma')}]$ requires $|\frac{\lambda+\lambda'}{2(1-\gamma\gamma')}|<1$. The condition $|\frac{\lambda+\lambda'}{2(1-\gamma\gamma')}|<1$ is equivalent to $|\frac{\gamma\gamma'+\lambda\lambda'}{1-\gamma\gamma'}|<1$. Together with the solution existence conditions $\gamma\gamma'(\gamma\gamma'-1)\neq0$ and $\Delta=0$ (see Eq. \eqref{solutions}), these constraints define the parameter space for DQPT-3, which can be summarized as
$\gamma\gamma'(\gamma\gamma'-1)\neq0$, $\Delta=0$ and $|\frac{\gamma\gamma'+\lambda\lambda'}{1-\gamma\gamma'}|<1$ (see Tab. \ref{Table}).

\textbf{Topological Properties of DQPT-3}

Considering $\gamma=1\rightarrow\gamma'=-0.5$ and $\lambda=1.1\rightarrow\lambda'=(1-2\gamma\gamma')\lambda-\sqrt{4\gamma\gamma'(1-\gamma\gamma')(1-\lambda^2)}$ as an example, we plot the Fisher zeros, the rate function (and its derivative), the winding number, and the trajectory in Fig. \ref{figure4}. The line of Fisher zeros touches the $\mathrm{Im}(z)$ axis at the critical mode [see Fig. \ref{figure4}(a)]. Although the rate function itself does not exhibit nonanalytic behavior (cusp singularity) [see the blue curve in Fig. \ref{figure4}(b)], its derivative does [see the green curve in Fig. \ref{figure4}(b)]. This singularity in $\mathrm{d}r(t)/\mathrm{d}t$ signals a qualitatively new type of DQPT. As $k$ varies from $0$ to $\pi$, the trajectory of $\vec{R}_k$ forms a closed loop that winds counterclockwise for $\lambda'>0$ [see Fig. \ref{figure4}(c)]. As time increases from $t=0$, this loop expands monotonically and eventually crosses the origin at the first critical time. Beyond this critical time, however, the loop moves away from the origin without ever enclosing it [see Fig. \ref{figure4}(c)]. When the trajectory passes through the origin at the critical time $t^{}$, the winding number exhibits a sudden jump from zero to a nonzero value, and immediately drops back to zero after $t^{}$. In other words, the transient jump in the winding number signals that the trajectory crosses the origin without encircling it, marking the occurrence of a DQPT-3.

\begin{figure*}
\begin{center}
\includegraphics[width=14cm]{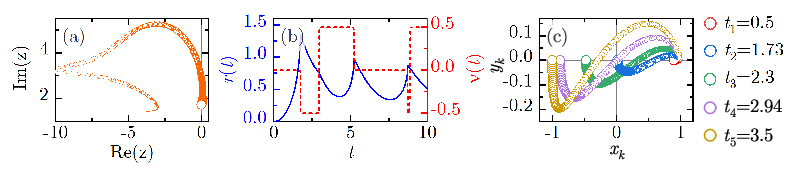}
\caption{(Color online) The properties of DQPT-5 are illustrated by the lines of Fisher zeros (a), the time evolution of rate function $r(t)$ (blue curves in b), the winding number $\nu(t)$ (red curves in b) and the trajectory of vector $\vec{R}_k=(x_k,y_k)$ (c) for the quenches from $\lambda=1$ to $\lambda'=0.1$. The first and the second critical times are $t_1^*\approx1.73$ and $t_2^*\approx2.94$. The other parameters are $\gamma=1$, $\gamma'=0.3$ and $n=0$.}
\label{figure6}
\end{center}
\end{figure*}

\begin{figure*}
\begin{center}
\includegraphics[width=14cm]{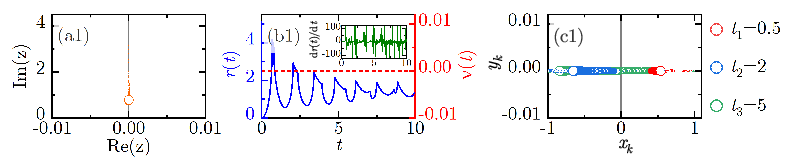}
\caption{(Color online) The properties of DQPT-6 are illustrated by the lines of Fisher zeros (a), the time evolution of rate function $r(t)$ (blue curves in b), the winding number $\nu(t)$ (red curves in b) and the trajectory of vector $\vec{R}_k=(x_k,y_k)$ (c) for the quenches from $\lambda=1$ to $\lambda'=-1$. The insert in (b) is the derivative of the rate function. The other parameters are $\gamma=0.5$, $\gamma'=2$ and $n=0$.}
\label{figure7}
\end{center}
\end{figure*}

\subsection{DQPT-4}

Here, we discuss the case of one critical boundary mode. For convenience, we denote the DQPT that occurs in this case as DQPT-4. Below, we discuss the parameter space and properties of DQPT-4 in detail, based on the solutions in Eq. \eqref{solutions}.

\textbf{Parameter Space of DQPT-4}

1. $\gamma\gamma'=1$:

A critical boundary mode can be found if $|\frac{1+\lambda\lambda'}{\lambda+\lambda'}|=1$, which requires $|\lambda|=1$ or $|\lambda'|=1$. Specifically, $k^*=\pi$ if $\lambda=1$ or $\lambda'=1$ and $k^*=0$ if $\lambda=-1$ or $\lambda'=-1$. If the transverse field is quenched to the quantum critical point $\lambda'=\pm1$, the critical boundary mode would not cause DQPT because $d_{k^*}'=0$ and the initial state does not evolve at all. The transverse field quenched from the quantum critical point $\lambda=\pm1$ to others $\lambda'\neq\lambda$ is a prerequisite of DQPT-4.

2. $\gamma\gamma'=0$:

According to Eq. \eqref{solutions}, a critical boundary mode $k^*=0$ or $\pi$ can be obtained if $\lambda=\mp1$ and $|\lambda'|>1$ or $|\lambda'|>1$ and $\lambda'=\mp1$. If the transverse field is quenched to the quantum critical point $\lambda'=\pm1$, then the critical boundary mode does not cause DQPT because $d_{k^*}'=0$ and the initial state does not evolve at all. A prerequisite for DQPT-4 is that the transverse field is quenched from the quantum critical point $\lambda=\pm1$ to another value $\lambda'\neq\lambda$. However, whether DQPT occurs also depends on the anisotropy parameter $\gamma$. If $\gamma=0$, the initial state becomes $|d_{k}^-\rangle=|0\rangle$ and the evolution operator at the critical boundary mode becomes $\hat{U}_{k^{*}}(t)=e^{-id_{k^{*}}'\sigma_zt}$, so that the initial state cannot be flipped by the evolution operator. Therefore, $\gamma\neq0$ is also a prerequisite for DQPT-4. If $\gamma\neq0$, the ground state (initial state) tends to the superposition $|d_{k^*}^-\rangle=\sqrt{1/2}(i|1\rangle+|0\rangle)$, while the evolution operator tends to $e^{-i(\lambda'\mp1)\hat{\sigma}^zt}$ as $k$ approaches the critical boundary mode $k^*$. In this case, the initial state can be flipped by the evolution operator, so DQPT-4 occurs. In summary, when $\gamma\gamma'=0$, DQPT-4 occurs under the following conditions: $\gamma\neq0$, $\gamma'=0$, $|\lambda|=1$, and $|\lambda'|>1$.

3. $\gamma\gamma'(\gamma\gamma'-1)\neq0$ and $\Delta>0$:

According to the discussions in DQPT-1, the solutions of Eq. \eqref{solution2} reduce to $\cos k_1=\mp1$ and $\cos k_2=\mp\frac{\lambda\lambda'+\gamma\gamma'}{1-\gamma\gamma'}$ when either $\lambda=\pm1$ or $\lambda'=\pm1$. Consequently, a critical boundary modes $k^*=\pi$ or $0$ can be found if $|\frac{\lambda\lambda'+\gamma\gamma'}{1-\gamma\gamma'}|>1$. If the transverse field is quenched to the quantum critical point $\lambda'=\pm1$, then the critical boundary mode does not cause DQPT because $d_{k^*}'=0$ and the initial state does not evolve at all. A prerequisite for DQPT-4 is that the transverse field is quenched from the quantum critical point $\lambda=\pm1$ to another value $\lambda'\neq\lambda$. In summary, when $\gamma\gamma'(\gamma\gamma'-1)\neq0$, DQPT-4 occurs under the following condition: $|\frac{\lambda\lambda'+\gamma\gamma'}{1-\gamma\gamma'}|>1$, $|\lambda|=1$ and $\lambda\neq\lambda'$.

All the elements consisting the parameter space of DQPT-4 are investigated above. To sum up, they consist of the following cases (see Tab. \ref{Table}):

(i) $\gamma\gamma'=1$, $|\lambda|=1$, $\lambda'\neq\lambda$;

(ii) $\gamma\neq0$, $\gamma'=0$, $|\lambda|=1$, $\lambda'\neq\lambda$;

(iii) $\gamma\gamma'(\gamma\gamma'-1)\neq0$, $|\lambda|=1$, $\lambda\neq\lambda'$, $\Delta>0$, $|\frac{\gamma\gamma'+\lambda\lambda'}{1-\gamma\gamma'}|>1$.

\textbf{Topological properties of DQPT-4}

Consider $\gamma=0.5\rightarrow\gamma'=2$ and $\lambda=1\rightarrow\lambda'=\pm1.5$ as an example, we plot the lines of Fisher zeros, the time evolution of rate function $r(t)$, the winding number $\nu(t)$ and the trajectory of vector of $\vec{R}_k=(x_k,y_k)$ in Fig. \ref{figure5}. It can be seen that the Fisher zeros coalesce into a continuous curve that touches $\mathrm{Im}(z)$ axis at a critical boundary mode $k^*$ [see Fig. \ref{figure5}(a1) and (a2)], giving rise to nonanalytic behavior (cusp singularity) of the rate function of the Loschmidt echo which implies DQPT occurring [see the blue curves in Fig. \ref{figure5}(b1) and (b2)]. It should be noted that the trajectory of $\vec{R}_k$ only swipes half a circle [see Fig. \ref{figure5}(c1) and (c2)]. This half-loop becomes larger and larger as time goes on from $t=0$ by moving its end, and its end  reaches the origin at the first critical time. After the critical time, the end of half-loop moves to the other side of the origin. In this case, the argument of the trajectory $\vec{R}_k$ (i.e., the geometric phase of the Loschmidt overlap amplitude) will suddenly changes with $\pi$, resulting in the jumping of the winding number with half-integer. This result is consistent with the observation in Ref. \cite{Ding2020,Wong2023}. The trajectory of vector $\vec{R}_k$ swipes half a circle anticounterclockwise for $\lambda'>0$ [see Fig. \ref{figure5}(c1)] and counterclockwise for $\lambda'<0$ [see Fig. \ref{figure5}(c2)], then the winding number jumps up [see the red curves in Fig. \ref{figure5}(b1)] or jumps down [see the red curves in Fig. \ref{figure5}(b2)] with half integer at the critical times.

\begin{figure*}
\begin{center}
\includegraphics[width=14cm]{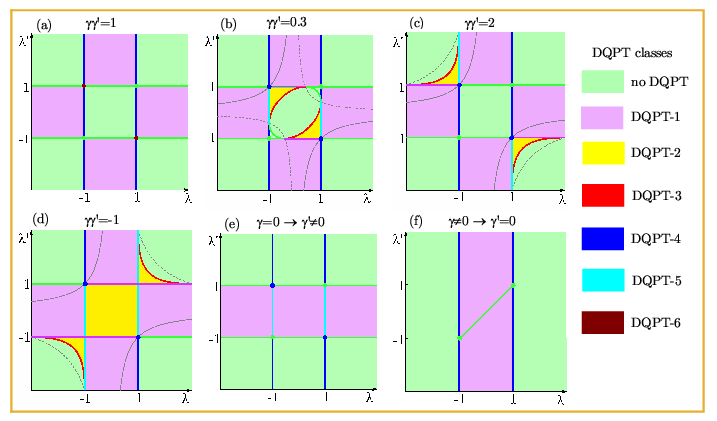}
\caption{(Color online) The dynamical phase diagrams. Note that the purple and green lines represent the same DQPTs as the purple and green areas. We have darkened the colors to emphasize the boundaries. }
\label{figure8}
\end{center}
\end{figure*}

\subsection{DQPT-5}

Now, we discuss the case of one critical boundary mode and one critical interior mode. For convenience, we denote the DQPT that occurs in this case as DQPT-5. Below, we discuss the parameter space and properties of DQPT-4 in detail, based on the solutions in Eq. \eqref{solutions} or in Eq. \eqref{solution2}.

\textbf{Parameter space of DQPT-5}

According to the discussions in DQPT-1, the solutions of Eq. \eqref{solution2} reduce to $\cos k_1=\mp1$ and $\cos k_2=\mp\frac{\lambda\lambda'+\gamma\gamma'}{1-\gamma\gamma'}$ when either $\lambda=\pm1$ or $\lambda'=\pm1$. Consequently, a critical boundary modes $k^*=\pi$ or $0$ and a critical interior mode $k^*_2=\arccos\bigl(\mp\frac{\lambda\lambda'+\gamma\gamma'}{1-\gamma\gamma'}\bigr)$ can be found if $|\frac{\lambda\lambda'+\gamma\gamma'}{1-\gamma\gamma'}|<1$. If the transverse field is quenched to the quantum critical point $\lambda'=\pm1$, then the critical boundary mode does not cause DQPT because $d_{k^*}'=0$ and the initial state does not evolve at all. A prerequisite for DQPT-4 is that the transverse field is quenched from the quantum critical point $\lambda=\pm1$ to another value $\lambda'\neq\lambda$. In summary, when $\gamma\gamma'(\gamma\gamma'-1)\neq0$, DQPT-5 occurs under the following condition: $|\frac{\lambda\lambda'+\gamma\gamma'}{1-\gamma\gamma'}|>1$, $|\lambda|=1$ and $\lambda\neq\lambda'$. Together with the solution existence conditions $\gamma\gamma'(\gamma\gamma'-1)\neq0$ and $\Delta>0$ (see Eq. \eqref{solutions}), these constraints define the parameter space for DQPT-5, which can be summarized as $\gamma\gamma'(\gamma\gamma'-1)\neq0$, $|\lambda|=1$, $\lambda\neq\lambda'$, $\Delta>0$ and $|\frac{\gamma\gamma'+\lambda\lambda'}{1-\gamma\gamma'}|<1$ (see Tab. \ref{Table}).

\textbf{Topological properties of DQPT-5}

Consider $\gamma=1\rightarrow\gamma'=0.3$ and $\lambda=1\rightarrow\lambda'=0.1$ as example, we plot DQPT-5 in Fig. \ref{figure6}. From Fig. \ref{figure6} we can see that the line of Fisher zeros cuts $\mathrm{Im}(z)$ axis at the critical mode $k_2^*$ and touches it at the critical mode $k_1^*$ [see Fig. \ref{figure6}(a)], giving rise to nonanalytic behavior of the rate function of the Loschmidt echo [see the blue curves in Fig. \ref{figure6}(b)]. As $k$ varies from $0$ to $\pi$, the coordinate of $\vec{R}_k$ moves first clockwise and then counterclockwise, forms an open trajectory [see Fig. \ref{figure6}(b)]. The end of trajectory crosses the origin at the first critical time, and then the winding number jumps down into $-1/2$ [see the red curves in Fig. \ref{figure6}(b)]. After the first critical time, the trajectory will half-encircle the origin until the second critical time, when the trajectory secondly cross the origin counterclockwise. Notably, it is not the end of trajectory that crosses the origin secondly. As a result, the winding number will jumps up from $-1/2$ to $1/2$ [see the red curves in Fig. \ref{figure6}(b)]. DQPT-5 is also an entirely new DQPT that is not discovered before.

\subsection{DQPT-6}
It is worth noting that when the transverse field is quenched from one critical point $\lambda=\pm1$ to another $\lambda'=\mp1$ and the anisotropic parameters satisfy $\gamma\gamma'=1$, Eq.~\eqref{critical mode} becomes an identity independent of $k$. In this case, all Fisher zeros lie on the $\mathrm{Im}(z)$ axis [see Fig. \ref{figure7}(a)], indicating a highly anomalous DQPT in which every instant is a critical time. That is, DQPTs occur at all times, with different critical modes responsible for the nonanalytic behavior at different moments. This DQPT is characterized by the rate function whose derivative exhibits singularities at all times, even though the rate function itself appears to show critical behavior only at some specific moments [see Fig. \ref{figure7}(b)]. At any given time, the trajectory of $\vec{R}_k$ lies entirely along the imaginary axis [see Fig. \ref{figure7}(c)] and invariably passes through the origin. Consequently, the winding number becomes ill-defined at every instant, as the geometric phase is not defined when the trajectory crosses the origin. This scenario stands in stark contrast to the DQPTs discussed previously, where the trajectory crosses the origin only at isolated critical times. In those cases, the winding number remains well-defined both before and after each crossing, enabling a topological classification of DQPTs. By contrast, the anomalous DQPT identified here eludes characterization by a winding number. This constitutes an entirely new type of DQPT, which we refer to as DQPT-6 in this paper.

In summary, we identify and characterize the critical interior mode $k^*\in(0,\pi)$ and the critical boundary mode $k^*=0$ or $\pi$ during the quench. The critical interior mode always induces DQPT with an integer-valued winding number, whereas the critical boundary mode always causes DQPT with a half-integer-valued winding number. By analysing the number and the nature of the critical modes, we provide a discussion of the topological properties of DQPTs. Based on their distinct topological features, we categorize DQPTs in one-dimensional XY model into six types:

\textbf{DQPT-1} is characterized by a winding number that jumps up or down by an integer step. It occurs when only one critical interior mode is generated and the line of Fisher zeros crosses the imaginary axis at this mode.

\textbf{DQPT-2} is characterized by a winding number alternates between upward and downward integer jumps. It occurs when two critical interior modes are present.

\textbf{DQPT-3} features a singularity in the winding number; that is, the winding number is always zero, and only at the moment when DQPT occurs does it suddenly jump by a finite value. DQPT-3 can also be described by the first derivative of the rate function. It occurs when Fisher zeros only touch the imaginary axis, where two critical interior modes merge into one by adjusting the parameters.

\textbf{DQPT-4} is characterized by a winding number that jumps up or down by a half-integer step. It occurs when a single critical boundary is present.

\textbf{DQPT-5} is characterized by alternating integer and half-integer jumps of the winding number. It occurs when a critical interior mode and a critical boundary mode appear simultaneously.

\textbf{DQPT-6} can not be described by the winding number. It occurs when all modes are critical, or equivalently, when all Fisher zeros lie on the imaginary axis.

Although DQPT-1, DQPT-2, and DQPT-4 have been previously reported, DQPT-3, DQPT-5, and DQPT-6 are entirely new and have not been discussed before. The conditions associated with each type of DQPT are summarized in Table \ref{Table}.

\section{Dynamical phase diagram}
Now we give the dynamical phase diagram according to the topological classification of DQPT and the associated conditions (see Table \ref{Table}). Although the conditions under which DQPTs occur are listed in Table 1, the dynamical phase diagram allows for a more clear and intuitive understanding.

We should consider six cases: $\gamma\gamma'<0$, $\gamma=0\rightarrow\gamma'\neq0$, $\gamma\neq0\rightarrow\gamma'=0$, $0<\gamma\gamma'<1$, $\gamma\gamma'=1$ and $\gamma\gamma'>1$ to investigate the dynamical phase diagrams, and plot them in Fig. \ref{figure8}. It can be seen that for $\gamma\gamma'\neq0$, i.e., the anisotropic parameters ($\gamma$ and $\gamma'$) before and after quench are both not at the critical point $\gamma_c=0$, DQPT-1 always happen if the transverse field is quenched across quantum critical point $|\lambda_c|=1$ [see the purple areas in Figs. \ref{figure8}(a)-(d)]. However, this does not mean DQPT must depend on the quantum phase transition because DQPT-2 (the yellow areas) and DQPT-3 (the red curves) can happen when the transverse field is quenched within the same phases. To be specific, DQPT-2 and DQPT-3 can happen at the ferromagnetic phase if $0<\gamma\gamma'<1$, but can happen at the paramagnetic phase if $\gamma\gamma'>1$ [see Fig. \ref{figure8}(b) and (c)]. Beyond that, DQPT-2 and DQPT-3 can happen at both ferromagnetic and paramagnetic phases, if the anisotropic parameter is quenched across the critical point $\gamma_c=0$ [see Fig. \ref{figure8}(d)]. It is worth noting that for a given suitable field $\lambda=\lambda'$, only the quench of the anisotropic parameter across the critical point $\gamma_c=0$ can still cause DQPT-2 and DQPT-3, and even DQPT-1.

At the critical boundaries, the situations are more complicated. If the transverse field is quenched from the quantum critical point $|\lambda_c|=1$, DQPT can always occur, either DQPT-4 or DQPT-5 (see the blue and cyan lines in Fig. \ref{figure8}), depending on $\lambda'$ and $\gamma\gamma'$, as shown in Table \ref{Table}. On the other hand, if the transverse field is quenched to the quantum critical point $|\lambda'|=1$, only DQPT-1 can occur, but not always, depending on $\lambda'$ and $\gamma\gamma'$ (see the purple lines in Fig. \ref{figure8}). If the anisotropic parameter is quenched from the critical point $\gamma_c=0$, DQPT-1 can always occur when the transverse field is quench to the ferromagnetic phase $|\lambda'|<1$, independent of the pre-quenched field $\lambda$ [see the purple areas in Fig. \ref{figure8}(e)]; on the contrary, if the anisotropic parameter is quenched to the critical point $\gamma_c=0$, DQPT-1 can always occur when the transverse field is quench from the ferromagnetically ordered phase $|\lambda'|<1$, independent of the post-quenched field $\lambda'\neq\lambda$ [see the purple areas in Fig. \ref{figure8}(f)]. These results furthermore demonstrate that DQPT is independent of quantum phase transition, which was first pointed out in Ref. \cite{Vajna2014}.

\section{Discussions}
It is worth emphasizing that the scope of our framework, which relies on the identification of critical interior and boundary modes, is not restricted to the XY model; it is applicable to other two-band models in one-dimensional systems, including the SSH model \cite{Li2014}, Rice-Mele model \cite{Lin2020}, Kitaev chain \cite{Vodola2014} and Creutz model \cite{Creutz1999}. While these models exhibit distinct Hamiltonian structures and belong to different symmetry classes \cite{McCann2023}, the orthogonality condition that signals the occurrence of DQPTs retains the same functional form as Eq.~\eqref{critical mode}. This underlying mathematical unity ensures the direct transferability of our theoretical framework to these diverse settings.

Here we consider SSH model as an example. Its Hamiltonian is
\begin{equation}\label{}
  H=\sum_nJ_1c_{n,A}^\dag c_{n,B}+J_2c_{n+1,A}^\dag c_{n,B}+h.c.,
\end{equation}
where $c_{n,\alpha}^\dag$ and $c_{n,\alpha}$ are the creation and annihilation operators on sublattice $\alpha\in \{A, B\}$ on the $n$th unit cell. $J_1$ and $J_2$ are the transition amplitudes of the intracell and intercell hopping processes, respectively. For this model,
\begin{equation}\label{}
  \mathbf{d}_k=(J_1+J_2\cos k,-J_2\sin k,0).
\end{equation}
According to the condition of $\mathbf{d}_k\cdot\mathbf{d}'_k=0$, we obtain
\begin{equation}\label{sshc}
  J_1J'_1+J_2J'_2+(J_1J'_2+J_2J'_1)\cos k=0.
\end{equation}
If $|J_1J'_2+J_2J'_1|>|J_1J'_1+J_2J'_2|$, then a critical interior mode $k^*=\arccos(-\frac{J_1J'_2+J_2J'_1}{J_1J'_2+J_2J'_1})$ can be found, implying that DQPT-1 will occur. In the contrary, if $|J_1|=|J_2|$ and $|J_1|\neq|J_2|$, a critical boundary mode can be found, which means that DQPT-4 can occur. At a very special case $J_1=J_2$, $J'_1=-J'_2$ or $J_1=-J_2$, $J'_1=J'_2$, Eq. (\ref{sshc}) is an equality independent of $k$. In other words, all the Fisher zeros are located on $\mathrm{Im}(z)$ axis, implying DQPT-6 occurs. In summary, only DQPT-1, DQPT-4 and DQPT-6 can occur in SSH model.

\section{Conclusions}

In this paper, we have investigated the topological properties of DQPTs in the one-dimensional XY model by identifying and characterizing critical interior modes and critical boundary modes during the quench. We have demonstrated that a critical interior mode always induces a DQPT with an integer-valued winding number, whereas a critical boundary mode always leads to a DQPT with a half-integer-valued winding number. By analyzing the number and classification of critical modes, we have provided a characterization of the topological properties of DQPTs in the one-dimensional XY model. Based on their distinct topological features, we have categorized DQPTs into six types:

\textbf{DQPT-1} is characterized by a winding number that jumps up or down by an integer step. It occurs when only one critical interior mode is generated and the line of Fisher zeros crosses the imaginary axis at this mode.

\textbf{DQPT-2} is characterized by a winding number that alternates between upward and downward integer jumps. It occurs when two critical interior modes are present.

\textbf{DQPT-3} features a singularity in the winding number; that is, the winding number is always zero, and only at the moment when DQPT occurs does it suddenly jump by a finite value. DQPT-3 can also be described by the first derivative of the rate function. It occurs when Fisher zeros only touch the imaginary axis, where two critical interior modes merge into one by adjusting the parameters.

\textbf{DQPT-4} is characterized by a winding number that jumps up or down by a half-integer step. It occurs when a single critical boundary mode is present.

\textbf{DQPT-5} is characterized by alternating integer and half-integer jumps of the winding number. It occurs when a critical interior mode and a critical boundary mode appear simultaneously.

\textbf{DQPT-6} cannot be described by the winding number. It occurs when all modes are critical, or equivalently, when all Fisher zeros lie on the imaginary axis.

Although DQPT-1, DQPT-2, and DQPT-4 have been previously reported in the literature, DQPT-3, DQPT-5, and DQPT-6 are entirely new and have not been discussed before. Furthermore, we have determined the parameter regimes for each type and have presented the corresponding phase diagrams. It is important to emphasize that our framework can be applied to other two-band models in one-dimensional systems, including the SSH model, Kitaev chain, Rice-Mele model, and Creutz model.

\appendix

\section{Derivation of the rate function and Fisher zeros}

In momentum space, the rate function can be expressed as
\begin{equation}\label{}
\begin{split}
r(t)&=-\lim_{N\rightarrow\infty}\frac{1}{N}\sum_{k>0}\ln|\mathcal{G}k(t)|^2 \\
&=-\lim_{N\rightarrow\infty}\frac{1}{N}\sum_{k>0}\ln\bigl[\mathcal{G}k(t)\mathcal{G}^*k(t)\bigr] \\
&=-\lim_{N\rightarrow\infty}\frac{2}{N}\sum_{k>0}\mathrm{Re}\bigl[\ln\mathcal{G}_k(t)\bigr].
\end{split}
\end{equation}
In the thermodynamic limit, the rate function takes the integral form
\begin{equation}\label{rt}
\begin{split}
r(t)=-\frac{1}{\pi}\mathrm{Re}\Bigl[\int_0^\pi\ln\mathcal{G}_k(t),dk\Bigr].
\end{split}
\end{equation}
Substituting the ground state, Eq.~\eqref{ground state}, and the time evolution operator, Eq.~\eqref{U}, the Loschmidt overlap amplitude $\mathcal{G}_k(t)$ can be written as
\begin{equation}\label{LOA}
\mathcal{G}_k(t)=\frac{1-\frac{\mathbf{d}_k}{d_k}\cdot\frac{\mathbf{d}_k'}{d_k'}}{2}e^{-id_k't}
+\frac{1+\frac{\mathbf{d}_k}{d_k}\cdot\frac{\mathbf{d}_k'}{d_k'}}{2}e^{id_k't}.
\end{equation}
Inserting this expression into Eq.~\eqref{rt} yields the rate function.

By replacing $it$ in Eq. \eqref{LOA} with $z$, we can obtain
\begin{equation}\label{}
\mathcal{G}_k(z)=\frac{1-\frac{\mathbf{d}_k}{d_k}\cdot\frac{\mathbf{d}_k'}{d_k'}}{2}e^{-zd_k'}
+\frac{1+\frac{\mathbf{d}_k}{d_k}\cdot\frac{\mathbf{d}_k'}{d_k'}}{2}e^{zd_k'}.
\end{equation}
Fisher zero in $\mathcal{G}_k(z)$ correspond to values of $z$ satisfying $\mathcal{G}_k(z) = 0$.
The condition $\mathcal{G}_k(z) = 0$ leads to
\begin{equation}\label{}
  e^{2zd_k'}=-\frac{1-\frac{\mathbf{d}_k}{d_k}\cdot\frac{\mathbf{d}_k'}{d_k'}}{1+\frac{\mathbf{d}_k}{d_k}\cdot\frac{\mathbf{d}_k'}{d_k'}}
\end{equation}
Solving this equation yields the Fisher zeros
\begin{equation}\label{Zeros}
z_n(k)=\frac{1}{2d_{k}'}\Biggl[\ln\frac{1-\frac{\mathbf{d}_k}{d_k}\cdot\frac{\mathbf{d}_k'}{d_k'}}
{1+\frac{\mathbf{d}_k}{d_k}\cdot\frac{\mathbf{d}_k'}{d_k'}}+i(2n+1)\pi\Biggr].
\end{equation}


\section*{References}
\begin{thebibliography}{}
\bibitem{Heyl2018} Heyl M 2018 Dynamical quantum phase transitions: A review \textit{Rep. Prog. Phys.} \textbf{81} 054001
\bibitem{Marino2022} Marino J, Eckstein M, Foster M S and Rey A M 2022 Dynamical phase transitions in the collisionless pre-thermal states of isolated quantum systems: Theory and experiments \textit{Rep. Prog. Phys.} \textbf{85} 116001

\bibitem{Bloch2008} Bloch I, Dalibard J and Zwerger W 2008 Many-body physics with ultracold gases \textit{Rev. Mod. Phys.} \textbf{80} 885
\bibitem{Greiner2002} Greiner M, Mandel O, H\"{a}nsch T W and Bloch I 2002 Quantum phase transition from a superfluid to a Mott insulator in a gas of ultracold atoms \textit{Nature} \textbf{419} 51
    
    
\bibitem{Porras2004} Porras D and Cirac J I 2004 Effective quantum spin systems with trapped ions \textit{Phys. Rev. Lett.} \textbf{92} 207901
\bibitem{Kim2009} Kim K, Chang M-S, Islam R, Korenblit S, Duan L-M and Monroe C 2009 Entanglement and tunable spin-spin couplings between trapped ions using multiple transverse modes \textit{Phys. Rev. Lett.} \textbf{103} 120502


\bibitem{Heyl2013} Heyl M, Polkovnikov A and Kehrein S 2013 Dynamical quantum phase transitions in the transverse field Ising model \textit{Phys. Rev. Lett.} \textbf{110} 135704
\bibitem{Heyl2015} Heyl M 2015 Scaling and universality at dynamical quantum phase transitions \textit{Phys. Rev. Lett.} \textbf{115} 140602

\bibitem{Schmitt2015} Schmitt M and Kehrein S 2015 Dynamical quantum phase transitions in the Kitaev honeycomb model \textit{Phys. Rev. B} \textbf{92} 075114
\bibitem{Rossi2022} Rossi L and Dolcini F 2022 Nonlinear current and dynamical quantum phase transitions in the flux-quenched Su-Schrieffer-Heeger model \textit{Phys. Rev. B} \textbf{106} 045410

\bibitem{Karrasch2013} Karrasch C and Schuricht D 2013 Dynamical phase transitions after quenches in nonintegrable models \textit{Phys. Rev. B} \textbf{87} 195104
\bibitem{Sharma2015} Sharma S, Suzuki S and Dutta A 2015 Quenches and dynamical phase transitions in a nonintegrable quantum Ising model \textit{Phys. Rev. B} \textbf{92} 104306

\bibitem{Kennes2018} Kennes D M, Schuricht D and Karrasch C 2018 Controlling dynamical quantum phase transitions \textit{Phys. Rev. B} \textbf{97} 184302
\bibitem{Halimeh2021a} Halimeh J C, Damme M V, Guo L, Lang J and Hauke P 2021 Dynamical phase transitions in quantum spin models with antiferromagnetic long-range interactions \textit{Phys. Rev. B} \textbf{104} 115133

\bibitem{Zhou2018} Zhou L, Wang Q-h, Wang H and Gong J 2018 Dynamical quantum phase transitions in non-Hermitian lattices \textit{Phys. Rev. A} \textbf{98} 022129
\bibitem{Zhou2021a} Zhou L and Du Q 2021 Non-Hermitian topological phases and dynamical quantum phase transitions: A generic connection \textit{New J. Phys.} \textbf{23} 063041

\bibitem{Mondal2022} Mondal D and Nag T 2022 Anomaly in the dynamical quantum phase transition in a non-Hermitian system with extended gapless phases \textit{Phys. Rev. B} \textbf{106} 054308
\bibitem{Mondal2023} Mondal D and Nag T 2023 Finite temperature dynamical quantum phase transition in a non-Hermitian system \textit{Phys. Rev. B} \textbf{107} 184311
\bibitem{Jing2024} Jing Y, Dong J-J, Zhang Y-Y and Hu Z-X 2024 Biorthogonal dynamical quantum phase transitions in non-Hermitian systems \textit{Phys. Rev. Lett.} \textbf{132} 220402


\bibitem{Bhattacharya2017a} Bhattacharya U, Bandyopadhyay S and Dutta A 2017 Mixed state dynamical quantum phase transitions \textit{Phys. Rev. B} \textbf{96}, 180303(R)
\bibitem{Heyl2017} Heyl M and Budich J C 2017 Dynamical topological quantum phase transitions for mixed states \textit{Phys. Rev. B} \textbf{96} 180304(R)
\bibitem{Lang2018} Lang H, Chen Y, Hong Q and Fan H 2018 Dynamical quantum phase transition for mixed states in open systems \textit{Phys. Rev. B} \textbf{98} 134310
\bibitem{Mera2018} Mera B, Vlachou C, Paunkovi\'{c} N, Vieira V R and Viyuela O 2018 Dynamical phase transitions at finite temperature from fidelity and interferometric Loschmidt echo induced metrics \textit{Phys. Rev. B} \textbf{97} 094110

\bibitem{Sharma2016} Sharma S, Divakaran U, Polkovnikov A and Dutta A 2016 Slow quenches in a quantum Ising chain: Dynamical phase transitions and topology \textit{Phys. Rev. B} \textbf{93} 144306
\bibitem{Maslowski2020} Mas{\l}owski T and Sedlmayr N 2020 Quasiperiodic dynamical quantum phase transitions in multiband topological insulators and connections with entanglement entropy and fidelity susceptibility \textit{Phys. Rev. B} \textbf{101} 014301
\bibitem{Zamani2024} Zamani S, Naji J, Jafari R and Langari A 2024 Scaling and universality at ramped quench dynamical quantum phase transitions \textit{J. Phys.: Condens. Matter} \textbf{36} 355401
\bibitem{Cheng2025} Cheng Y, Zhang Y, Qiu T, Xin P, and Xu B-M 2025 Dynamic Phase Transitions in Periodically Driving 1D Ising Model \textit{arXiv}:2512.24600
\bibitem{Baghran2024} Baghran R, Jafari R and Langari A 2024 Competition of long-range interactions and noise at a ramped quench dynamical quantum phase transition: The case of the long-range pairing Kitaev chain \textit{Phys. Rev. B} \textbf{110} 064302

\bibitem{Yang2019} Yang K, Zhou L, Ma W, Kong X, Wang P, Qin X, Rong X, Wang Y, Shi F, Gong J and Du J 2019 Floquet dynamical quantum phase transitions \textit{Phys. Rev. B} \textbf{100} 085308
\bibitem{Zamani2020} Zamani S, Jafari R and Langari A 2020 Floquet dynamical quantum phase transition in the extended XY model: Nonadiabatic to adiabatic topological transition \textit{Phys. Rev. B} \textbf{102} 144306

\bibitem{Jafari2022} Jafari R, Akbari A, Mishra U and Johannesson H 2022 Floquet dynamical quantum phase transitions under synchronized periodic driving \textit{Phys. Rev. B} \textbf{105} 094311
\bibitem{Naji2022} Naji J, Jafari R, Zhou L and Langari A 2022 Engineering Floquet dynamical quantum phase transitions \textit{Phys. Rev. B} \textbf{106} 094314
\bibitem{Jafari2021} Jafari R and Akbari A 2021 Floquet dynamical phase transition and entanglement spectrum \textit{Phys. Rev. A} \textbf{103} 012204    
\bibitem{Trapin2021} Trapin D, Halimeh J C and Heyl M 2021 Unconventional critical exponents at dynamical quantum phase transitions in a random ising chain \textit{Phys. Rev. B} \textbf{104} 115159

\bibitem{Kuliashov2023} Kuliashov O N, Markov A A and Rubtsov A N 2023 Dynamical quantum phase transition without an order parameter \textit{Phys. Rev. B} \textbf{107} 094304

\bibitem{Jafari2024} Jafari R, Langari A, Eggert S and Johannesson H 2024 Dynamical quantum phase transitions following a noisy quench \textit{Phys. Rev. B} \textbf{109} L180303

\bibitem{Cao2020} Cao K, Li W, Zhong M and Tong P 2020 Influence of weak disorder on the dynamical quantum phase transitions in the anisotropic XY chain \textit{Phys. Rev. B} \textbf{102} 014207



\bibitem{Zhang2025} Zhang X, Hu L, and Li F 2025 Universal scaling in Loschmidt echo across quantumphase transitions under nonadiabatic dynamics, \textit{Phys. Rev. B} \textbf{112} 024310


\bibitem{Jurcevic2017} Jurcevic P, Shen H, Hauke P, Maier C, Brydges T, Hempel C, Lanyon B P, Heyl M, Blatt R and Roos C F 2017 Direct observation of dynamical quantum phase transitions in an interacting many-body system \textit{Phys. Rev. Lett.} \textbf{119} 080501
\bibitem{Flaschner2018} Fl\"{a}schner N, Vogel D, Tarnowski M, Rem B S, L\"{u}hmann D-S, Heyl M, Budich J C, Mathey L, Sengstock K and Weitenberg C 2018 Observation of dynamical vortices after quenches in a system with topology \textit{Nat. Phys.} \textbf{14} 265
\bibitem{Guo2019} Guo X-Y, Yang C, Zeng Y, Peng Y, Li H-K, Deng H, Jin Y-R, Chen S, Zheng D and Fan H 2019 Observation of a dynamical quantum phase transition by a superconducting qubit simulation \textit{Phys. Rev. Applied} \textbf{11} 044080

\bibitem{Budich2016} Budich J C and Heyl M 2016 Dynamical topological order parameters far from equilibrium \textit{Phys. Rev. B} \textbf{93} 085416
\bibitem{Bhattacharya2017b} Bhattacharya U and Dutta A 2017 Emergent topology and dynamical quantum phase transitions in two-dimensional closed quantum systems \textit{Phys. Rev. B} \textbf{96}, 014302
\bibitem{Kosior2024} Kosior A and Heyl M 2024 Vortex loop dynamics and dynamical quantum phase transitions in three-dimensional fermion matter \textit{Phys. Rev. B} \textbf{109} L140303

\bibitem{Ding2020} Ding C 2020 Dynamical quantum phase transition from a critical quantum quench \textit{Phys. Rev. B} \textbf{102} 060409(R)
\bibitem{Qiu2018} Qiu X, Deng T-S, Guo G-C and Yi W 2018 Dynamical topological invariants and reduced rate functions for dynamical quantum phase transitions in two dimensions \textit{Phys. Rev. A} \textbf{98} 021601(R)

\bibitem{Wong2023} Wong C Y, Cheraghi H and Yu W C 2023 Quantum spin fluctuations in dynamical quantum phase transitions \textit{Phys. Rev. B} \textbf{108} 064305

\bibitem{Bunder1999} Bunder J E and McKenzie R H 1999 Effect of disorder on quantum phase transitions in anisotropic XY spin chains in a transverse field \textit{Phys. Rev. B} \textbf{60} 344

\bibitem{Fisher1965} Fisher M E 1965 in Boulder Lectures in Theoretical Physics (University of Colorado, Boulder) Vol. 7.

\bibitem{Cheraghi2024} Cheraghi H, Sirker J, and Sedlmayr N 2024 Fisher zeroes and dynamical quantum phase transitions for two- and three-dimensional models \textit{Phys. Rev. B} \textbf{110} 224302

\bibitem{Vajna2014} Vajna S and D\'{o}ra B 2014 Disentangling dynamical phase transitions from equilibrium phase transitions \textit{Phys. Rev. B} \textbf{89} 161105(R)

\bibitem{Vajna2015} Vajna S and D\'{o}ra B 2015 Topological classification of dynamical phase transitions \textit{Phys. Rev. B} \textbf{91} 155127

\bibitem{Porta2020} Porta S, Cavaliere F, Sassetti M, and Ziani N T 2020 Topological classification of dynamical quantum phase transitions in the XY chain \textit{Sci. Rep.} \textbf{10} 12766

\bibitem{Xu2024} Xu B-M 2024 Quantum coherence assisted dynamical phase transition \textit{Commun. Theor. Phys.} \textbf{76} 125104


\bibitem{Li2014} Li L, Xu Z, and Chen S 2014 Topological phases of generalized Su-Schrieffer-Heeger models \textit{Phys. Rev. B} \textbf{89} 085111
\bibitem{Lin2020} Lin Y-T, Kennes D M, Pletyukhov M, Weber C S, Schoeller H, and Meden V 2020 Interacting Rice-Mele model: Bulk and boundaries \textit{Phys. Rev. B} \textbf{102} 085122
\bibitem{Vodola2014} Vodola D, Lepori L, Ercolessi E, Gorshkov A V and Pupillo G 2014 Kitaev chains with long-range pairing \textit{Phys. Rev. Lett.} \textbf{113} 156402
\bibitem{Creutz1999} Creutz M 1999 End states, ladder compounds, and domain-wall fermions \textit{Phys. Rev. Lett.} \textbf{83} 2636

\bibitem{McCann2023}  McCann E 2023 Catalog of noninteracting tight-binding models with two energy bands in one dimension \textit{Phys. Rev. B} \textbf{107} 245401














\end{thebibliography}
\end{document}